\documentclass[%
reprint,
 amsmath,amssymb,
]{revtex4-1}

\usepackage{bm}
\usepackage{graphicx}
\usepackage{amsfonts}
\usepackage{amscd}
\usepackage{enumerate}
\usepackage{subfigure}
\usepackage{amsthm}
\usepackage{dcolumn}
\usepackage{amsmath,amssymb}

\usepackage{txfonts}

\usepackage[breaklinks]{hyperref}

\begin{document}
\title{Passive decoy-state quantum key distribution for the weak coherent photon source with intensity fluctuations}
\author{Yuan Li}
\author{Wan-su Bao}
\email{2010thzz@sina.com}
\author{Hong-wei Li}
\author{Chun Zhou}
\author{Yang Wang}
\affiliation{Zhengzhou Information Science and Technology Institute, Zhengzhou 450004, China}

\begin{abstract}

Passive decoy-state quantum key distribution (QKD) systems, proven to be more desirable than active ones in some scenarios, also have the problem of device imperfections like intensity fluctuations. In this paper, the formula of key generation rate of the passive decoy-state protocol using transformed weak coherent pulse (WCP) source with intensity fluctuation is given, and then the influence of intensity fluctuations on the performance of passive decoy-state protocol is rigorously characterized. From numerical simulations, it can be seen that intensity fluctuations have non-negligible influence on the performance of the passive decoy-state QKD protocol with WCP source. Most importantly, our simulations show that, under the same deviation of intensity fluctuations, the passive decoy-state method performs better than the active two-intensity decoy-state method and is close to the active three-intensity decoy-state method.

\end{abstract}

\maketitle

\section{introduction}

Quantum key distribution (QKD)\cite{Bennett1984,Ekert1991}, allows two legitimate parties, Alice and Bob, to create a random secret key even when the channel is accessible to an eavesdropper, Eve. Compared with classical cryptography, quantum cryptography has the biggest advantage that its unconditionally security is based on the fundamental laws of physics---no-cloning theorem and uncertainty principle\cite{Nielsen2000}.

Since the best-known protocol--BB84\cite{Bennett1984} was proposed by Bennett and Brassard, quantum cryptography has developed well both theoretically and experimentally\cite{Wang2012,Jouguet2013,Ma2012,Wang2013a,Lo2012,Braunstein2012,Zhou2013,Wang2013b}. In the original proposal of BB84 protocol, a single photon source is necessary. But the single photon source is still commercially unavailable with current technology. Usually people use weak coherent pulse (WCP) source instead and many WCP-based QKD experiments have been done since the first QKD experiment\cite{Bennett1992}. Actually due to the multi-photon pulse, the QKD system will suffer from photon-number splitting (PNS) attack\cite{Huttner1995,Brassard2000}. To protect QKD from PNS attack, one can use the so-called decoy-state method\cite{Hwang2003,Lo2005,Wang2005a,Wang2005b,Ma2005} that could closely reach the performance of single photon sources. The method that Alice prepares decoy state actively is also called active decoy-state method. But in practice, because of the imperfect experiments and channels, it may bring in some side channel information that Eve can make use of to have an attack. ¡°In real active (regular) decoy state experiments, it is more difficult to verify the assumption that Eve cannot distinguish decoy and signal states.¡±\cite{Ma2008}

However, passive decoy-state method\cite{Mauere2007,AYKI2007} can reduce the side information in decoy state preparation procedure. Different from the active decoy-state method, passive decoy-state method only uses one intensity signal, and it passively distinguishes decoy and signal states by Alice's detector. Passive decoy-state method doesn't need to change the experiments that active decoy-state method has used.

Existing studies of decoy-state method always suppose that the devices are ideal, and Alice, Bob can control their experiments accurately. In fact, the conditions are difficult to satisfy, especially for the practical photon sources. An important imperfect factor of photon sources is intensity fluctuations. Due to unavoidable interference from environments, the power of source is constantly changing. And there should be deviation between the true value and assumed value. When the deviation rises and falls irregularly, one can call it intensity fluctuation. The intensity fluctuation in experiment will result in the irregular change of the photon number distribution, bringing potential security threat to the practical QKD. Therefore, researching on the impact of intensity fluctuation to the security of QKD system is far important to the practical application of QKD.

For active decoy-state method, Wang et al.\cite{Wang2008,Wang2009} have studied the relationship between key generation rate and source errors. And WCP source is regarded as a specific example to prove the correctness of the conclusions. Wang et al., Zhou\cite{WangS2009,Zhou2010} have proven that HSPS and HPCS sources are more stable than WCP sources in the conditions of intensity fluctuation, respectively. However, most of the above results are all confined within the active decoy-state method. J. Z. Hu et al.\cite{JZ2010} have a significative study on the AYKI protocol about the PDC source with intensity fluctuations. The WCP source used in passive decoy-state method also exist the imperfection of intensity fluctuation. Therefore, how intensity fluctuations influence the performance of passive decoy-state method that uses WCP source remains to be studied and this is just the motivation of this paper.

In this paper, we firstly introduce the transformed WCP source used for passive decoy state. With the source, we describe the passive decoy-state protocol that without considering intensity fluctuations. Then we consider the passive decoy-state method using the transformed WCP source with intensity fluctuation. And we recalculate the final key rate of passive decoy-state method with intensity fluctuations. By numerical simulations, we give the results of key generation rate with different transmission distances and different intensity fluctuations. Finally we compare the passive decoy-state method with the active decoy-state method of three-intensity and two-intensity.

The paper is organized as follows. In Sec. \ref{S2} we introduce the transformed WCP source. Next in Sec. \ref{S3} we study the passive decoy-state method using transformed WCP source in the ideal condition. Then we consider the passive decoy-state method using transformed WCP source with intensity fluctuations in Sec. \ref{S4}. The numerical simulations of Sec. \ref{S3} and \ref{S4} are shown in Sec. \ref{S5}. And we also show the comparison between the passive decoy-state method with the active decoy-state method of three-intensity and two-intensity in this Sec. Finally, Sec. \ref{S6} concludes the paper with a summary.

\section{The transformed WCP source}\label{S2}

Due to the characters of WCP source itself, it cannot be used for passive decoy-state method directly. Curty et al.\cite{Curty2009,Curty2010} transform WCP source to make the source output two Fock diagonal states, so that it can be used for passive decoy-state. The fundamental setups are shown as follows.

\begin{figure}[ht!]
\centering
\includegraphics[width=80mm]{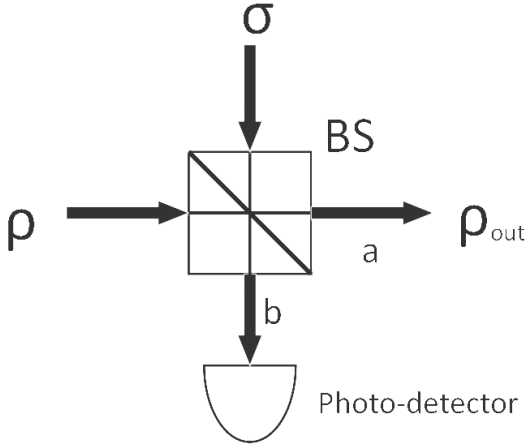}
\caption{\label{fig1}Transformed WCP sources.}
\end{figure}

In Fig.1, $\rho$ and $\sigma$ denote the coherent states of two phase randomized WCP source states, respectively,
\begin{equation}\label{E1}
\rho  = {e^{ - {\mu _1}}}\sum\limits_{n = 0}^\infty  {\frac{{\mu _1^n}}{{n!}}} \left| n \right\rangle \left\langle n \right|,
\end{equation}
\begin{equation}\label{E2}
\sigma  = {e^{ - {\mu _2}}}\sum\limits_{n = 0}^\infty  {\frac{{\mu _2^n}}{{n!}}} \left| n \right\rangle \left\langle n \right|,
\end{equation}
with $\mu_1$ and $\mu_2$ denoting the mean photon number of the two signals, respectively. In this paper, we consider the threshold detector as the Photon-detector in Fig.1. In this scenario, the joint probability of having $n$ photons in output mode $a$ and $m$ photons in output mode $b$ can be written as\cite{Curty2009}
\begin{equation}\label{E3}
{p_{n,m}} = \frac{{{\upsilon ^{n + m}}{e^{ - \upsilon }}}}{{n!m!}}\frac{1}{{2\pi }}{\int_0^{2\pi } {{\gamma ^n}(1 - \gamma )} ^m}d\theta,
\end{equation}
where the parameters $\upsilon$, $\gamma$ and $\xi$ are given by
\begin{equation}\label{E4}
\begin{array}{l}
 \upsilon  = {\mu _1} + {\mu _2}, \\
 \gamma  = \frac{{{\mu _1}t + {\mu _2}(1 - t) + \xi \cos \theta }}{\upsilon }, \\
 \xi  = 2\sqrt {{\mu _1}{\mu _2}(1 - t)t} . \\
 \end{array}
\end{equation}
and $t$ denote the transmittance of a beam splitter.

Whenever the sender, Alice, does not care the result of the measurement in mode $b$, the probability of having n photons in mode a can be written as
\begin{equation}\label{E5}
p_n^t = \sum\limits_{m = 0}^\infty  {{p_{n,m}}}  = \frac{{{\upsilon ^n}}}{{n!}}\frac{1}{{2\pi }}\int_0^{2\pi } {{\gamma ^n}{e^{ - \upsilon \gamma }}} d\theta.
\end{equation}

For the Alice's detector, the joint probability of having $n$ photons in mode $a$ and no click in the threshold detector has now the form can be expressed as
\begin{equation}\label{E6}
\begin{array}{l}
 p_n^{\overline c } = \left( {1 - \varepsilon } \right)\sum\limits_{m = 0}^\infty  {{{\left( {1 - {\eta _d}} \right)}^m}{p_{n,m}}}  \\
  = \left( {1 - \varepsilon } \right)\frac{{{\upsilon ^n}{e^{ - {\eta _d}\upsilon }}}}{{n!}}\frac{1}{{2\pi }}{\int_0^{2\pi } {{\gamma ^n}e} ^{ - \left( {1 - {\eta _d}} \right)\upsilon \gamma }}d\theta . \\
 \end{array}
\end{equation}
where the parameter $\varepsilon$ denotes dark count and $\eta_d$ denotes the detection efficiency of the detector.

Then the probability for having $n$ photons in mode $a$ and producing a click in Alice¡¯s threshold detector is
\begin{equation}\label{E6}
p_n^c = p_n^t - p_n^{\bar c}.
\end{equation}

\section{Passive decoy-state method using transformed WCP source without intensity fluctuations}\label{S3}

The difference between passive decoy state and active decoy state is that passive decoy state need not change the intensity of the laser pulses randomly to estimate the lower bound of single-photon counts ${Q_1}$ and the upper bound of quantum bit error rate (QBER) of bits generated by single photon pulses. In passive decoy state, the key point to distinguish decoy states and signal states is click or no click of Alice's detector.

The common point between passive decoy state and active decoy state is that the counting rates and the error rates of pulse of the same photon-number states from the signal states and the decoy states shall be equal to each other\cite{Lo2005},
\begin{equation}\label{E7}
{Y_m} = Y'_m,{e_m} = e'_m,
\end{equation}
where ${Y_m}$ and $Y'_m$ denote the counting rates of m photons from signal states and decoy states, respectively. And ${e_m}$, $e'_m$ denote the error rates of $m$ photons from signal states and decoy states,  respectively.

In 2005, Lo et al.\cite{Lo2005} combined the decoy state method with the results provided by Gottesman-Lo-L\"utkenhaus-Preskill(GLLP)\cite{Gottesman2004} analysis and gave the exact formula for secure key generation rate:
\begin{equation}\label{E8}
R \ge \sum\limits_{l = c,\bar c} {\max \left\{ {{R^l},0} \right\}} ,
\end{equation}
where $R^l$ satisfies
\begin{equation}\label{E9}
{R^l} \ge q\left\{ {p_0^l{Y_0} + p_1^l{Y_1}\left[ {1 - {H_2}\left( {{e_1}} \right)} \right] - {Q^l}f\left( {{E^l}} \right){H_2}\left( {{E^l}} \right)} \right\}.
\end{equation}

The parameter $q$ is the efficiency of the protocol. For the standard BB84 protocol\cite{Bennett1984}, $q=1/2$. $f\left( {{E^l}} \right)$ denotes the efficiency of the error correction protocol as a function of the error rate ${E^l}$\cite{Brassard1994}, typically $f\left( {{E^l}} \right) \ge 1$ with Shannon limit $f\left( {{E^l}} \right) = 1$(usually, we consider 1.22 as its approximate value). ${e_1}$ denotes the single photon error rate. ${H_2}\left( x \right) =  - x{\log _2}\left( x \right) - \left( {1 - x} \right){\log _2}\left( {1 - x} \right)$ is the binary Shannon entropy function.

For passive decoy state, it has said that one distinguishes decoy states and signal states by click and no click of Alice's detector. So in this paper we define that the $c$ denotes ¡°click¡± of the detector while $\bar c$ denotes ¡°no click¡± of the detector. And here we denote the signals that cause a click of  Alice's detector are signal states. The ones that cause no click of Alice's detector are decoy states.

Then the $l$ in Eq.\ref{E9} is $l \in \left( {c,\bar c} \right)$.
And ${Q_c}$ , ${Q_{\bar c}}$ denote the counting rates of Alice¡¯s detector producing a click and no click, respectively,
\begin{equation}\label{E10}
{Q_c} = 1 - (1 - {Y_0}){e^{ - {\eta _d}w}}{I_{0,{\eta _{sys}}\xi }},
\end{equation}
\begin{equation}\label{E11}
{Q_{\bar c}} = {N_w} - (1 - \varepsilon )(1 - {Y_0}){e^{\left( {{\eta _d} - {\eta _{sys}}} \right)w - {\eta _d}v}}.
\end{equation}
where ${I_{q,z}}$ represents the modified Bessel function of the first kind\cite{Arfken1985}.

The upper bound of ${e_1}$ and the lower bound of $Y_1^L$ is\cite{Curty2009,Curty2010}
\begin{eqnarray}\label{E12}
\begin{array}{l}
 e_1^U = \min \left\{ {\frac{{{E_c}{Q_c} - p_0^cY_0^L{e_0}}}{{p_1^cY_1^L}}} \right.,\frac{{{E_{\bar c}}{Q_{\bar c}} - p_0^{\bar c}Y_0^L{e_0}}}{{p_1^{\bar c}Y_1^L}}, \\
 \\~~~~~~~~~~~~~~~~~~~~~
 \left. {\frac{{p_0^{\bar c}{E_t}{Q_t} - p_0^t{E_{\bar c}}{Q_{\bar c}}}}{{\left( {p_1^tp_0^{\bar c} - p_1^{\bar c}p_0^t} \right)Y_1^L}}} \right\}, \\
 \end{array}
\end{eqnarray}
\begin{equation}\label{E13}
Y_1^L = \max \left\{ {\frac{{p_2^t{Q_{\bar c}} - p_2^{\bar c}{Q_t} - \left( {p_2^tp_0^{\bar c} - p_2^{\bar c}p_0^t} \right){Y_0}}}{{p_2^tp_1^{\bar c} - p_2^{\bar c}p_1^t}},0} \right\},
\end{equation}
where
\begin{equation}\label{E14}
{E_t} = {{\left( {{e_0} - {e_d}} \right){Y_0}} \mathord{\left/
 {\vphantom {{\left( {{e_0} - {e_d}} \right){Y_0}} {{Q^t}}}} \right.
 \kern-\nulldelimiterspace} {{Q^t}}} + {e_d},
\end{equation}
\begin{equation}\label{E15}
{E_{\bar c}} = {{\left( {{e_0} - {e_d}} \right){Y_0}} \mathord{\left/
 {\vphantom {{\left( {{e_0} - {e_d}} \right){Y_0}} {{N_w}}}} \right.
 \kern-\nulldelimiterspace} {{N_w}}} + {e_d},
\end{equation}
\begin{equation}\label{E16}
{E_c} = {E_t} - {E_{\bar c}},
\end{equation}
\begin{equation}\label{E17}
{N_w} = \sum\limits_{n = 0}^\infty  {p_n^{\bar c}}  = (1 - \varepsilon ){e^{ - {\eta _d}\left[ {{\mu _1}\left( {1 - t} \right) + {\mu _2}t} \right]}}{I_{0,{\eta _d}\xi }}.
\end{equation}
The final result shall be shown in Sec.\ref{S5}.

\section{Passive decoy-state method using transformed WCP source with intensity fluctuations}\label{S4}
Now we introduce the parameter $\delta$ that denotes the intensity fluctuations. The fluctuation ranges of the two intensities of the WCP states are characterized by
\begin{equation}\label{E18}
{\mu_1}\left( {1 - {\delta _{{\mu_1}}}} \right) \le \mu_1^{rea} \le {\mu_1}\left( {1 + {\delta _{{\mu_1}}}} \right),
\end{equation}
\begin{equation}\label{E19}
{\mu_2}\left( {1 - {\delta _{{\mu_2}}}} \right) \le \mu_2^{rea} \le {\mu_2}\left( {1 + {\delta _{{\mu_2}}}} \right),
\end{equation}
where $\mu_1^{rea}$ and $\mu_2^{rea}$ are the real intensity of WCP states $\rho$ and $\sigma$. The range of the intensity fluctuation parameters ${\delta _{{\mu_1}}}$ and ${\delta _{{\mu_2}}}$ is $\left[ {0,0.10} \right]$. When ${\delta _x} = 0(x = {\mu_1},{\mu_2})$, it means there are no intensity fluctuations. When ${\delta _x} = 0.10(x = {\mu_1},{\mu_2})$, it signifies the maximum upper bound of the intensity fluctuation.
According to the range of $\mu_1^{rea}$ and $\mu_2^{rea}$, we can get
\begin{equation}\label{E20}
p_c^L \le p_c^{rea} \le p_c^U,
p_{\bar c}^L \le p_{\bar c}^{rea} \le p_{\bar c}^U,
\end{equation}
where $p_c^{rea}$ and $p_{\bar c}^{rea}$ are the real probability for having n photons in mode a in the case of having a or no click in the threshold detector, respectively.

Now we suppose Alice totally sends $M$ optical pulses to Bob. The count of Alice¡¯s detector ¡°click¡± is ${N_c}$ and the count of Alice¡¯s detector ¡°no click¡± is ${N_{\bar c}}$. ${N_c}$ and ${N_{\bar c}}$ can be expressed as
\begin{equation}\label{E21}
{N_c} = \sum\limits_{m = 0}^\infty  {{n_{m,c}}}  = {n_{0,c}} + {n_{1,c}} + {n_{2,c}} + \sum\limits_{m = 3}^\infty  {{n_{m,c}}},
\end{equation}
\begin{equation}\label{E22}
{N_{\bar c}} = \sum\limits_{m = 0}^\infty  {{n_{m,\bar c}}}  = {n_{0,\bar c}} + {n_{1,\bar c}} + {n_{2,\bar c}} + \sum\limits_{m = 3}^\infty  {{n_{m,\bar c}}},
\end{equation}
where ${n_{m,c}}$ and ${n_{m,\bar c}}$ denote counts of having $m$ photons in mode $a$ when Alice¡¯s detector are click and no click, respectively. The formulas can also be written as
\begin{equation}\label{E23}
{N_c} = {P_c}M{Q_c},
{N_{\bar c}} = {P_{\bar c}}M{Q_{\bar c}},
\end{equation}
where ${P_c}$ denotes the probability of Alice¡¯s detector producing a click, ${P_{\bar c}}$ denotes the probability of Alice¡¯s detector producing no click,
\begin{equation}\label{E24}
{P_{\bar c}} = {N_w} = \sum\limits_{n = 0}^\infty  {p_n^{\bar c}}  = (1 - \varepsilon ){e^{ - {\eta _d}\left[ {{\mu _1}\left( {1 - t} \right) + {\mu _2}t} \right]}}{I_{0,{\eta _d}\xi }},
\end{equation}

The fraction of m-photon counts is\cite{Wang2007}
\begin{equation}\label{E25}
{\Delta _{m,x}} = \frac{{{Y_{m,x}}{p_{m,x}}}}{{{Q_x}}} = \frac{{{n_{m,x}}}}{{{N_x}}}.\left( {x = c,\bar c} \right)
\end{equation}

Our purpose is to estimate the lower bound of ${\Delta _{1,c}}$ and ${\Delta _{0,c}}$ in the intensity fluctuation case.

In order to cancel ${n_{m \ge 2,c}}$  and ${n_{m \ge 2,\bar c}}$  elements in Eq.\ref{E21} \ref{E22}, applying $\frac{{{N_{\bar c}}}}{{{P_{\bar c}}}} - \frac{{{N_c}}}{{{P_c}}}$, we can get
\begin{equation}\label{E26}
\begin{array}{l}
 \frac{{{n_{1,\bar c}}}}{{{P_{\bar c}}}} - \frac{{{n_{1,c}}}}{{{P_c}}} = \left( {\frac{{{N_{\bar c}}}}{{{P_{\bar c}}}} - \frac{{{N_c}}}{{{P_c}}}} \right) + \left( {\frac{{{n_{0,c}}}}{{{P_c}}} - \frac{{{n_{0,\bar c}}}}{{{P_{\bar c}}}}} \right) \\
  + \sum\limits_{m = 2}^\infty  {\left( {\frac{{{n_{m,c}}}}{{{P_c}}} - \frac{{{n_{m,\bar c}}}}{{{P_{\bar c}}}}} \right)} . \\
 \end{array}
\end{equation}

And we also note that the following inequalities hold true:\cite{Ma2005,Ma2006}
\begin{equation}\label{E27}
0 \le {n_{0,c}} \le \frac{{{E_c}{N_c}}}{{{e_0}}},{n_{0,\bar c}} \\
\le \frac{{{P_{\bar c}}{n_{0,c}}}}{{{P_c}{q_0}}},
\end{equation}
\begin{equation}\label{E28}
{n_{m,\bar c}} \le \frac{{{P_v}p_{m,v}^U}}{{{P_\mu }p_{m,\mu }^L}}{n_{m,c}} = \frac{{{P_v}}}{{{P_\mu }}}\frac{{{n_{m,c}}}}{{{q_m}}}
\end{equation}
where ${q_m} = {{p_{m,\mu }^L} \mathord{\left/{\vphantom {{p_{m,\mu }^L} {p_{m,v}^U}}} \right.\kern-\nulldelimiterspace} {p_{m,v}^U}}$ and $k\ge 0$.

By substituting inequalities Eq.\ref{E27} and Eq.\ref{E28}, the third and forth elements in Eq.\ref{E26} become
\begin{equation}\label{E29}
\frac{{{n_{0,c}}}}{{{P_c}}} - \frac{{{n_{0,\bar c}}}}{{{P_{\bar c}}}} \ge  - \frac{{{E_c}{N_c}}}{{{e_0}}}\left( {\frac{{{1}}}{{{q_0}}} - 1} \right),
\end{equation}
\begin{equation}\label{E30}
\frac{{{n_{m,c}}}}{{{P_c}}} - \frac{{{n_{m,\bar c}}}}{{{P_{\bar c}}}} \ge  - \frac{{{n_{m,c}}}}{{{P_c}}}\left( {1 - \frac{{{1}}}{{{q_m}}}} \right).
\end{equation}

And when $m \ge 1$ we can get ${q_m} \le {q_1}$ and ${q_2} < {q_0}$ . Finally we obtain the lower bound of the fraction of single-photon count for the signal source,
\begin{equation}\label{E31}
{\Delta _{1,c}} = \frac{{{n_{1,c}}}}{{{N_c}}} \ge \Delta _{1,c}^L = \frac{{{Q_{\bar c}} - {Q_c} - \left( {\frac{{{1}}}{{{q_0}}} - 1} \right)\frac{{{E_c}{N_c}}}{{{e_0}}}}}{{\left( {\frac{{{1}}}{{{q_0}}} - 1} \right){Q_c}}}.
\end{equation}

Then the lower bound of ${\Delta _{0,c}}$ is
\begin{equation}\label{E32}
{\Delta _{0,c}} = \frac{{{n_{0,\mu }}}}{{{N_\mu }}} \ge \Delta _{0,c}^L = \frac{{{Q_0}p_{0,\mu }^L}}{{{Q_\mu }}}.
\end{equation}

We define ${e_{m,x}}\left( {x = c,\bar c} \right)$ is the error rate of m photons state when Alice¡¯s detector produces a or no click, respectively. The overall quantum bit error rate (QBER) is
\begin{equation}\label{E33}
{E_x} = \sum\limits_{m = 0}^\infty  {{e_{m,x}}{\Delta _{m,x}}} ,\left( {x = c,\bar c} \right).
\end{equation}

Since ${e_{m,x}} \ge 0$ and ${\Delta _{m,x}} \ge 0$ for all $m \ge 2$, we can formulate the upper bound of the single-photon error rate of Alice's detector producing a click,
\begin{equation}\label{E34}
{e_{1,c}} \le e_{1,c}^U = \frac{{{E_c} - {e_0}\Delta _{0,c}^L}}{{\Delta _{1,c}^L}} = \frac{{{E_c}}}{{\Delta _{1,c}^L}},
\end{equation}
where the error rate of dark count ${e_0}$ is 0.5.

Then we can get the key rate generated by the GLLP formula\cite{Gottesman2004}.

Now we shall calculate ${R_{\bar c}}$. In Sec.\ref{S3} we say that counting rates and error rates of pulse of the same photon-number states from the signal states and the decoy states must be equal.
\begin{equation}\label{E35}
{Y_{m,c}} = {Y_{m,\bar c}},{e_{m,c}} = {e_{m,\bar c}}.
\end{equation}

According to Eq.\ref{E25} and Eq.\ref{E31}, we can get
\begin{equation}\label{E36}
Y_{1,\bar c}^l = Y_{1,c}^l = \frac{{\Delta _{1,c}^l{Q_c}}}{{p_{1,c}^u}}.
\end{equation}

Then, the lower bound of $\Delta _{1,\bar c}^l$ can be expressed as
\begin{equation}\label{E37}
\Delta _{1,\bar c}^l = \frac{{Y_{1,\bar c}^lp_{1,\bar c}^u}}{{{Q_{\bar c}}}}.
\end{equation}

After all, the key generation rate ${R^{\bar c}}$ is
\begin{equation}\label{E38}
{R^{\bar c}} \ge q{Q_{\bar c}}\left\{ {\Delta _{0,\bar c}^L + \Delta _{1,\bar c}^L\left[ {1 - {H_2}\left( {e_{1,\bar c}^U} \right)} \right] - f\left( {{E_{\bar c}}} \right){H_2}\left( {{E_{\bar c}}} \right)} \right\}.
\end{equation}

And the final key rate generated is
\begin{equation}\label{E39}
R = {R^c} + {R^{\bar c}}.
\end{equation}

\section{Numerical simulations}\label{S5}

Here we will take some numerical simulations to show how intensity fluctuation influences on final key generation rate.

In this scenario, the yields ${Y_n}$ can be expressed as\cite{Lo2005,Ma2005}
\begin{equation}
{Y_n} = 1 - \left( {1 - {Y_0}} \right){\left( {1 - {\eta _{sys}}} \right)^n},
\end{equation}
where ${\eta _{sys}}$ denotes the overall transmittance of the system. It can be written as
\begin{equation}
{\eta _{sys}} = {\eta _{channel}}{\eta _{Bob}},
\end{equation}
where ${\eta _{channel}}$ is the transmittance of the quantum channel, and ${\eta _{Bob}}$ denotes the overall transmittance of Bob¡¯s detection apparatus. The parameter ${\eta _{channel}}$ can be related with a transmission distance $d$ measured in km for the given QKD schemes as
\begin{equation}
{\eta _{channel}} = {10^{ - \frac{{\alpha d}}{{10}}}},
\end{equation}
where $\alpha$ represents the loss coefficient of the channel (e.g., an optical fiber) measured in dB/km.

Firstly, we shall describe the final key generation rate of passive decoy state using WCP source without intensity fluctuation. The experimental QKD parameters showed in Table I come from\cite{Gobby2004}.

\begin{table}[!b]
\begin{tabular}{c c c c c}
\hline\hline  ${\alpha(dB/km)}$  &  ${\eta_{Bob}}$  &  ${e_d}$  &  ${Y_0}$  &  ${\eta_d}$  \\
\hline  $0.21$  &  $0.045$  &  $0.033$  &  $1.7\times {{10}^{-6}}$  &  $0.12$  \\
\hline\hline
\end{tabular}
\caption {Experimental QKD parameters}
\end{table}

We set as \cite{Curty2009,Curty2010} mentioned that ${\mu _1}$ is around 0.5, ${\mu _2}$ is quite weak (around $10^{-4}$). Also we set $\varepsilon$ is $3.2\times {{10}^{-7}}$. Define $R$ as the key generation rate of passive decoy state. Define $R(\delta)/R(0)$ is the fidelity of passive decoy state, where $R\left( \delta  \right)$ denotes the $R$ with intensity fluctuations and $R\left( 0 \right)$ denotes the R with no intensity fluctuations.

Now we will characterize the relationship between R and the intensity fluctuation when transmission distance $d$ is fixed. The result is shown in Fig. 2.
\begin{figure}[ht!]
\centering
\includegraphics[width=80mm]{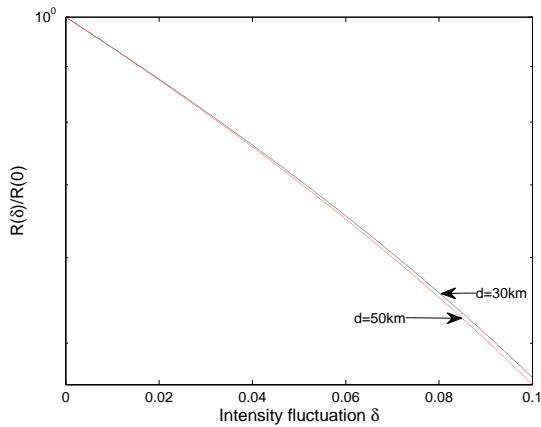}
\caption{\label{Fig2} (Color online) R versus intensity fluctuations $\delta$.}
\end{figure}

From Fig.2, we can see that the $R(\delta)/R(0)$ of $d=30km$ is larger than the one of $d=50km$. And when $\delta$ is getting to 0.1, the $R(\delta)/R(0)$ is getting to 0. It indicates that the key generation rate $R$ is becoming small with intensity fluctuation becoming large.

Then Fig.3 and Fig.4 shows us the comparison of the key rate $R$ versus transmission distance among two-intensity, three-intensity decoy-state method and passive decoy-state method.
\begin{figure}[ht!]
\centering
\includegraphics[width=80mm]{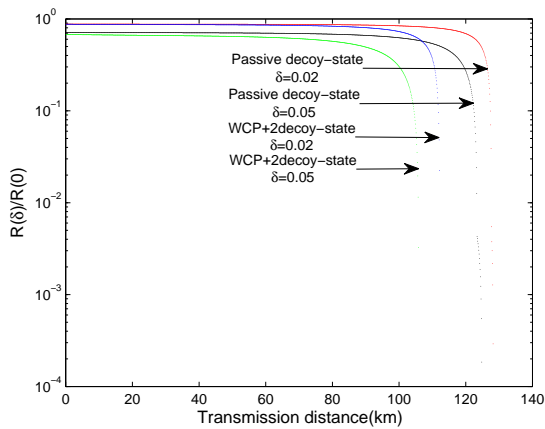}
\caption{(Color online) Comparison of the $R(\delta)/R(0)$ versus transmission distance with intensity fluctuations between passive decoy-state and WCP+2-decoy-state.}
\end{figure}

\begin{figure}[ht!]
\centering
\includegraphics[width=80mm]{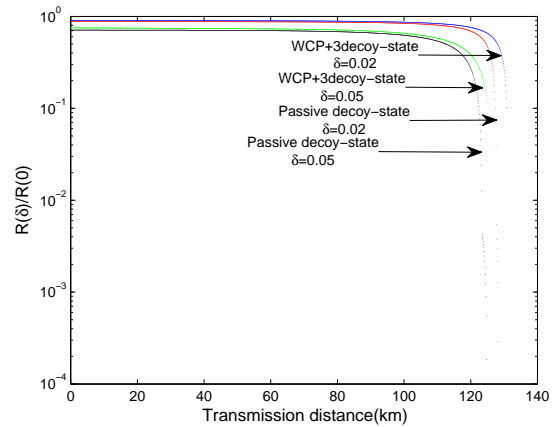}
\caption{(Color online) Comparison of the $R(\delta)/R(0)$ versus transmission distance with intensity fluctuations between passive decoy-state and WCP+3-decoy-state.}
\end{figure}

In Fig.3 and Fig.4, the $R(\delta)/R(0)$ and the extreme transmission distance of passive decoy-state method with $\delta=0.02$ are larger than those of $\delta=0.05$, obviously. We can get that $R$ becomes smaller with intensity fluctuation $\delta$ becoming larger for passive decoy-state method. So is the extreme transmission distance. Also, we can find that the passive decoy-state method is better than the two-intensity active decoy-state method and is close to the three-intensity active decoy-state method. The $R(\delta)/R(0)$ of passive decoy-state method is almost equal to the one of three-intensity method with $\delta=0.02$. The different point is that the extreme transmission distance of passive decoy-state method is a little smaller than that of the three-intensity method. We analyze the reasons as follows. Firstly, the passive decoy-state method only uses the sets of click and no click. This may lead to reduction of key generation efficiency but is still close to the theoretical limit. Secondly, it has been proven that the three-intensity method approaches the infinite decoy-state method\cite{Hwang2003}. And the passive decoy-state method can be regard as a two-intensity passive decoy-state method. So the transmission distance may be a little smaller that the three-intensity method, the ideal case. But considering the factors of the implementation of the three-intensity active decoy-state scheme and the protection from the side channel information attacks that may bring in the possibility of distinguishing decoy and signal states, the passive decoy-state method is better.

From the simulation results above, we can easily see that intensity fluctuations have unignorable influence on the final key rate of the passive decoy-state QKD protocol with WCP source. When the transmission distance is close to its upper bound, the influence is obvious particularly.

\section{Conclusions}\label{S6}

In summary, we analyze the transformed WCP source that can be use for passive decoy states. Using the source, we recalculated the final key rate of passive decoy-state method with intensity fluctuations. According to the numerical simulations, we find that the intensity fluctuations have influence that cannot be neglected on the final key rate of the passive decoy-state QKD method with WCP source. When the intensity fluctuation parameter $\delta$ becomes large, the key generation rate and the extreme transmission distance become small. Moreover, comparing with two-intensity, three-intensity active decoy-state method, we can get that the passive decoy-state is better than two-intensity active decoy-state and is close to three-intensity active decoy-state in the case of having intensity fluctuations. But when we consider the difficulty of implementation and the side channel information which may bring in the risk of distinguishing decoy and signal states by Eve, passive decoy-state method is a better choice.



\end{document}